\documentclass[preprint,prd,aps,showpacs,showkeys,nofootinbib]{revtex4}
\usepackage{amsmath}
\usepackage{amsmath,graphicx}
\begin{document}

\title{\( B_s^0 \rightarrow \mu^+ \mu^- \) in a flavor violating extension of MSSM}

\author{Kai-Kai Meng$^{1,2,}$\footnote{mkkpaper@foxmail.com},
Hai-Bin Zhang$^{1,2,3,4,}$\footnote{hbzhang@hbu.edu.cn}, Jin-Lei Yang$^{1,2,3}$\footnote{jlyang@hbu.edn.cn}
}

\affiliation{
$^1$Department of Physics, Hebei University, Baoding, 071002, China\\
$^2$Hebei Key Laboratory of High-precision Computation and Application of
Quantum Field Theory, Baoding, 071002, China\\
$^3$Hebei Research Center of the Basic Discipline for Computational Physics, Baoding, 071002, China\\
$^4$Institute of Life Science and Green Development, Hebei University, Baoding, 071002, China
}

\begin{abstract}		
$B$ meson rare decays play a crucial role in exploring new physics beyond the standard model. In this study, we explore the rare decay process $B_s^0 \rightarrow \mu^+ \mu^-$ in a flavor violating extension of the Minimal Supersymmetric Standard Model (MSSM), namely the $\mu$-from-$\nu$ SSM ($\mu\nu$SSM). Combined with the decay $\bar{B}\rightarrow X_s\gamma$, the numerical results indicate that the $\mu\nu$SSM can successfully accommodate the experimental data for $B_s^0 \rightarrow \mu^+ \mu^-$ and additionally narrow down the parameter space.
\end{abstract}

\keywords{Rare decay, $\mu\nu$SSM, B physics}
\pacs{12.60.Jv, 14.80.Da}

\maketitle

\section{Introduction\label{sec1}}
\indent\indent
While the Standard Model (SM) has achieved great success in describing known phenomena, it is still believed to require improvement or expansion to describe physics at higher energy scales. Beyond the SM, supersymmetry is regarded as one of the most credible candidates. In supersymmetric(SUSY) theory, novel TeV-scale particles can feature in competing diagrams, inducing detectable effects on the rate or other attributes of the \( b \to s \) decay process. The examinations of rare $B$ decays offer a means to identify new physics beyond the SM as they are less susceptible to uncertainties stemming from nonperturbative QCD effects. Recently, the average experimental data on the branching ratios of \( \bar{B} \to X_s \gamma \) and \( B_s^0 \to \mu^+ \mu^- \) are reported as follows~\cite{HFLAV:2022esi,ParticleDataGroup:2024cfk,ATLAS:2018cur,CMS:2019bbr,LHCb:2021vsc,CMS:2022mgd}
\begin{eqnarray}
	&&Br(\bar B\rightarrow X_s \gamma)=(3.49\pm0.19)\times 10^{-4},\nonumber\\
	&&Br(B_s^0\rightarrow \mu^+\mu^-)=(3.34\pm0.27)\times10^{-9}.
	\label{experimental data}
\end{eqnarray}
The SM predicts the branching ratios for \( \bar B \to X_s \gamma \) recently as~\cite{Misiak:2020vlo}
\begin{eqnarray}
	&&Br(\bar B\rightarrow X_s \gamma)=(3.40\pm0.17)\times 10^{-4}.
	\label{SM bsr}
\end{eqnarray}
The SM prediction for \( B_s^0 \to \mu^+ \mu^- \) including power-enhanced QED correction is~\cite{Beneke:2017vpq,Beneke:2019slt,Czaja:2024the}
\begin{eqnarray}
	&&Br(B_s^0\rightarrow \mu^+\mu^-)=(3.64\pm0.12)\times10^{-9}.
	\label{SM Buu}
\end{eqnarray}

These SM predictions align well with the experimental results, indicating that the precise measurements of rare $B$ decay processes impose stringent constraints on new physics beyond the SM. The primary objective of investigating $B$ decays is to seek evidence of new physics and define its parameter space.

Indeed, the analyses of constraints on extensions of the SM are widely discussed in the literature. The calculation of the rate inclusive decay \( \bar{B} \to X_s \gamma \) is detailed by the authors of Refs.~\cite{Ciuchini1,Ciafaloni,Borzumati1} within the framework of the Two-Higgs doublet model (THDM). The impact of supersymmetry on \(\bar{B} \to X_s \gamma \) is deliberated in Refs.~\cite{NPB4,NPB5,NPB6,NPB8,NPB7,Zhang1,Feng1}, while the next-to-leading order (NLO) QCD corrections are provided in Ref.~\cite{NPB9}. The branching ratio for \( B^0_s \to l^+ l^- \) in the THDM and SUSY extensions of the SM has been calculated in Refs.~\cite{He:1988tf,Skiba:1992mg,Choudhury:1998ze,Huang:2000sm,Feng2,Feng3}. Additionally, Ref.~\cite{NPB11} delves into hadronic $B$ decays, while CP-violation in these processes is discussed in Ref.~\cite{NPB12}. Ref.~\cite{NPB13} explores the potential for observing SUSY effects in rare decays \( \bar{B} \to X_s \gamma \) and \( B \to X_s l^+ l^- \) at the B-factory. The investigation of SUSY effects on these processes is highly intriguing, and research on them could illuminate the fundamental features of the SUSY model. The pertinent reviews can be found in Refs.~\cite{NPB16,NPB17}.

In the context of the Supersymmetric Standard Model with a neutrino Yukawa sector ($\mu\nu$SSM)\cite{ref2,ref3,ref4} ,the model addresses the $\mu$ problem \cite{ref5} that arises in the MSSM \cite{ref6,ref7,ref8}. This resolution is facilitated through the inclusion of lepton number-breaking couplings between the right-handed neutrino superfields and the Higgs fields \( \epsilon_{ab} \lambda_i \tilde{\nu}_i^c \tilde{H}_d^a \tilde{H}_u^b \) in the superpotential. The $\mu$ term is spontaneously generated through the vacuum expectation values (VEVs) of the right-handed neutrino superfields, denoted as $\mu = \lambda_i \langle \tilde{\nu}_i^c \rangle$, upon the breaking of the electroweak symmetry (EWSB).

In our previous work, we have investigated the decay $\bar{B}\rightarrow X_s\gamma$ in the $\mu\nu$SSM~\cite{Zhang1}.
In this paper, we investigate the flavor changing neutral current (FCNC) processes $B_s^0\rightarrow \mu^+\mu^-$ within the framework of the $\mu\nu$SSM using a minimal flavor-violating scenario for the soft breaking terms, combined with the decay $\bar{B}\rightarrow X_s\gamma$. Our presentation is structured as follows: Section II provides a brief summary of the key components of the $\mu\nu$SSM, encompassing the superpotential and general soft breaking terms. Section III presents the effective Hamiltonian for $B_s^0 \to \mu^+ \mu^-$. The numerical analyses are detailed in Section IV, with Section V offering a summary. Appendix contains the detailed formulas.

\section{the $\mu\nu$SSM\label{sec2}}
\indent\indent
Compared to the MSSM, the \( \mu\nu \mathrm{SSM} \) includes three right-handed sneutrino superfields \( \widetilde{\nu}_i^c \) (where \( i = 1, 2, 3 \)) with non-zero VEVs. The superpotential of the \( \mu\nu \mathrm{SSM} \) can be expressed as~\cite{ref2}:
\begin{eqnarray}
	&&W ={\epsilon _{ab}} ({Y_{{u_{ij}}}}\hat H_u^b\hat Q_i^a\hat u_j^c + {Y_{{d_{ij}}}}\hat H_d^a\hat Q_i^b\hat d_j^c
	+ {Y_{{e_{ij}}}}\hat H_d^a\hat L_i^b\hat e_j^c \nonumber \\
	&&\qquad + {Y_{{\nu _{ij}}}}\hat H_u^b\hat L_i^a\hat \nu _j^c ) -  {\epsilon _{ab}}{\lambda _i}\hat \nu _i^c\hat H_d^a\hat H_u^b + \frac{1}{3}{\kappa _{ijk}}\hat \nu _i^c\hat \nu _j^c\hat \nu _k^c\;.
	\label{super-w}
\end{eqnarray}
where \( \epsilon_{ab} \) represents the antisymmetric tensor. \( \hat{H}_d^T = \left( \hat{H}_d^0, \hat{H}_d^- \right) \), \( \hat{H}_u^T = \left( \hat{H}_u^+, \hat{H}_u^0 \right) \), \( \hat{Q}_i^T = \left( \hat{u}_i, \hat{d}_i \right) \), and \( \hat{L}_i^T = \left( \hat{\nu}_i, \hat{e}_i \right) \) represent $SU(2)$ doublet superfields. The symbols \( \hat{d}_j^c \), \( \hat{u}_j^c \), and \( \hat{e}_j^c \) denote the singlet superfields corresponding to the down-type quark, up-type quark, and lepton, respectively. Furthermore, \( Y \), \( \lambda \), and \( \kappa \) are dimensionless matrices, a vector, and a totally symmetric tensor. The indices \( a \), \( b = 1, 2 \) are $SU(2)$ indices, while \( i \), \( j \), \( k = 1, 2, 3 \) represent generation indices.

In Eq.~(\ref{super-w}), the initial three terms mirror those found in the MSSM. Following the EWSB, the subsequent terms can generate effective bilinear expressions such as \( \epsilon_{ab} \varepsilon_i \hat{H}_b^u \hat{L}_a^i \) and \( \epsilon_{ab} \mu \hat{H}_d^a \hat{H}_b^u \), where \( \varepsilon_i = Y_{\nu_{ij}} \langle \tilde{\nu}_j^c \rangle \) and \( \mu = \lambda_i \langle \tilde{\nu}_i^c \rangle \). The final two terms explicitly break lepton number and R-parity, with the last term capable of generating effective Majorana masses for neutrinos at the electroweak scale. Throughout this paper, the summation convention is assumed for repeated indices.

In the $\mu\nu$SSM, the general soft SUSY-breaking terms are as follows:
\begin{eqnarray}
	&&- \mathcal{L}_{soft}\:=\:m_{{{\tilde Q}_{ij}}}^{\rm{2}}\tilde Q{_i^{a\ast}}\tilde Q_j^a
	+ m_{\tilde u_{ij}^c}^{\rm{2}}\tilde u{_i^{c\ast}}\tilde u_j^c + m_{\tilde d_{ij}^c}^2\tilde d{_i^{c\ast}}\tilde d_j^c
	+ m_{{{\tilde L}_{ij}}}^2\tilde L_i^{a\ast}\tilde L_j^a  \nonumber\\
	&&\hspace{1.8cm} + \: m_{\tilde e_{ij}^c}^2\tilde e{_i^{c\ast}}\tilde e_j^c + m_{{H_d}}^{\rm{2}} H_d^{a\ast} H_d^a
	+ m_{{H_u}}^2H{_u^{a\ast}}H_u^a + m_{\tilde \nu_{ij}^c}^2\tilde \nu{_i^{c\ast}}\tilde \nu_j^c \nonumber\\
	&&\hspace{1.8cm}  + \: \epsilon_{ab}{\left[{{({A_u}{Y_u})}_{ij}}H_u^b\tilde Q_i^a\tilde u_j^c
		+ {{({A_d}{Y_d})}_{ij}}H_d^a\tilde Q_i^b\tilde d_j^c + {{({A_e}{Y_e})}_{ij}}H_d^a\tilde L_i^b\tilde e_j^c + {\rm{H.c.}} \right]} \nonumber\\
	&&\hspace{1.8cm}  + \left[ {\epsilon _{ab}}{{({A_\nu}{Y_\nu})}_{ij}}H_u^b\tilde L_i^a\tilde \nu_j^c
	- {\epsilon _{ab}}{{({A_\lambda }\lambda )}_i}\tilde \nu_i^c H_d^a H_u^b
	+ \frac{1}{3}{{({A_\kappa }\kappa )}_{ijk}}\tilde \nu_i^c\tilde \nu_j^c\tilde \nu_k^c + {\rm{H.c.}} \right] \nonumber\\
	&&\hspace{1.8cm}  - \: \frac{1}{2}\left({M_3}{{\tilde \lambda }_3}{{\tilde \lambda }_3}
	+ {M_2}{{\tilde \lambda }_2}{{\tilde \lambda }_2} + {M_1}{{\tilde \lambda }_1}{{\tilde \lambda }_1} + {\rm{H.c.}} \right).\label{soft-term}
\end{eqnarray}
Here, the first two lines feature mass-squared terms of squarks, sleptons, and Higgs bosons. The following two lines encompass the trilinear scalar couplings. The final line specifies the Majorana masses for gauginos \( \tilde {\lambda}_3 \), \( \tilde {\lambda}_2 \), and \( \tilde {\lambda}_1 \) denoted as $M_3$, $M_2$, and $M_1$ respectively. Besides the terms from \( \mathcal{L}_{\text{soft}} \), the tree-level scalar potential also receives the typical contributions from $D$ and $F$ terms~\cite{ref3}.

After spontaneous breaking of the electroweak symmetry, the neutral scalars typically acquire VEVs:
\begin{eqnarray}
	\langle H_d^0 \rangle = \upsilon_d , \qquad \langle H_u^0 \rangle = \upsilon_u , \qquad
	\langle \tilde \nu_i \rangle = \upsilon_{\nu_i} , \qquad \langle \tilde \nu_i^c \rangle = \upsilon_{\nu_i^c}.\label{VEVs}
\end{eqnarray}
One has the option to characterize the neutral scalars as
\begin{eqnarray}
	&&H_d^0=\frac{h_d + i P_d}{\sqrt{2}} + \upsilon_d, \qquad\; \tilde \nu_i = \frac{(\tilde \nu_i)^\Re + i (\tilde \nu_i)^\Im}{\sqrt{2}} + \upsilon_{\nu_i},  \nonumber\\
	&&H_u^0=\frac{h_u + i P_u}{\sqrt{2}} + \upsilon_u, \qquad \tilde \nu_i^c = \frac{(\tilde \nu_i^c)^\Re + i (\tilde \nu_i^c)^\Im}{\sqrt{2}} + \upsilon_{\nu_i^c},
\end{eqnarray}
and
\begin{eqnarray}
	\tan\beta={\upsilon_u\over\sqrt{\upsilon_d^2+\upsilon_{\nu_i}\upsilon_{\nu_i}}}.
\end{eqnarray}
Then, given that \( \upsilon_{\nu_i} \ll v_d, v_u \), we can define the value of \( \tan \beta \) as usual, where \( \tan \beta = \frac{v_u}{v_d} \).

In the \(8 \times 8\) charged scalar mass matrix \(M^2_{S^{\pm}}\), there exist the massless unphysical Goldstone bosons \(G^{\pm}\) which can be expressed as~\cite{ref-zhang,ref15,ref-zhang-LFV,ref-zhang1}:
\begin{eqnarray}
	G^{\pm} = {1 \over \sqrt{\upsilon_d^2+\upsilon_u^2+\upsilon_{\nu_i} \upsilon_{\nu_i}}} \Big(\upsilon_d H_d^{\pm} - \upsilon_u {H_u^{\pm}}-\upsilon_{\nu_i}\tilde e_{L_i}^{\pm}\Big).
\end{eqnarray}
In the unitary gauge, these Goldstone bosons \(G^{\pm}\) are absorbed by the $W$-boson and are no longer present in the Lagrangian. Consequently, the mass squared of the $W$-boson is given by:
\begin{eqnarray}
	m_W^2={e^2\over2s_{_W}^2}\Big(\upsilon_u^2+\upsilon_d^2+\upsilon_{\nu_i} \upsilon_{\nu_i}\Big),
\end{eqnarray}
where \(e\) represents the electromagnetic coupling constant, \(s_{W} = \sin\theta_{W}\) with \(\theta_{W}\) being the Weinberg angle.

\section{Theoretical calculation on $Br(B_s^0\rightarrow \mu^+\mu^-)$\label{sec3}}
\indent\indent
The effective Hamiltonian for the $b\to s \mu^+\mu^-$ transition at the hadronic scale can be expressed as~\cite{bobeth,bobeth02}:
\begin{equation} \label{eq:Heff}
	{\cal H}_{eff} = - \frac{4\,G_F}{\sqrt{2}}\left(
	\lambda_t {\cal H}_{eff}^{(t)} + \lambda_u {\cal
		H}_{eff}^{(u)}\right),
\end{equation}
with the CKM combination $\lambda_i=V_{ib}V_{is}^*$ and
\begin{eqnarray}
	{\cal H}_{eff}^{(t)}
	& = &
	C_1 \mathcal O_1^c + C_2 \mathcal O_2^c + \sum_{i=3}^{6} C_i
	\mathcal O_i + \sum_{i=7,8,9,10,P,S} (C_i \mathcal O_i + C'_i \mathcal
	O'_i)\,,
	\nonumber\\
	{\cal H}_{eff}^{(u)}
	& = &
	C_1 (\mathcal O_1^c-\mathcal O_1^u)  + C_2(\mathcal O_2^c-\mathcal
	O_2^u)\,.
\end{eqnarray}

\begin{figure}
	\setlength{\unitlength}{1mm}
	\centering
	\includegraphics[width=5.5in]{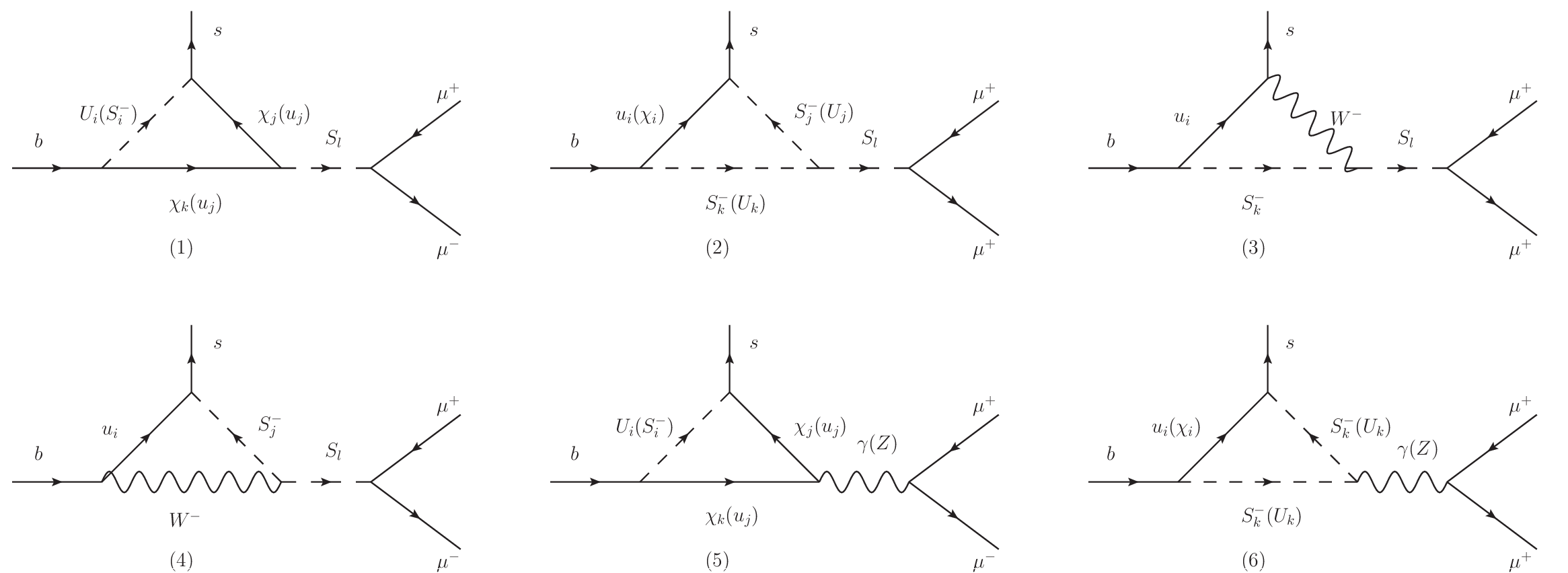}
	\vspace{0cm}
	\par
	\hspace{-0.in}
	\includegraphics[width=5.5in]{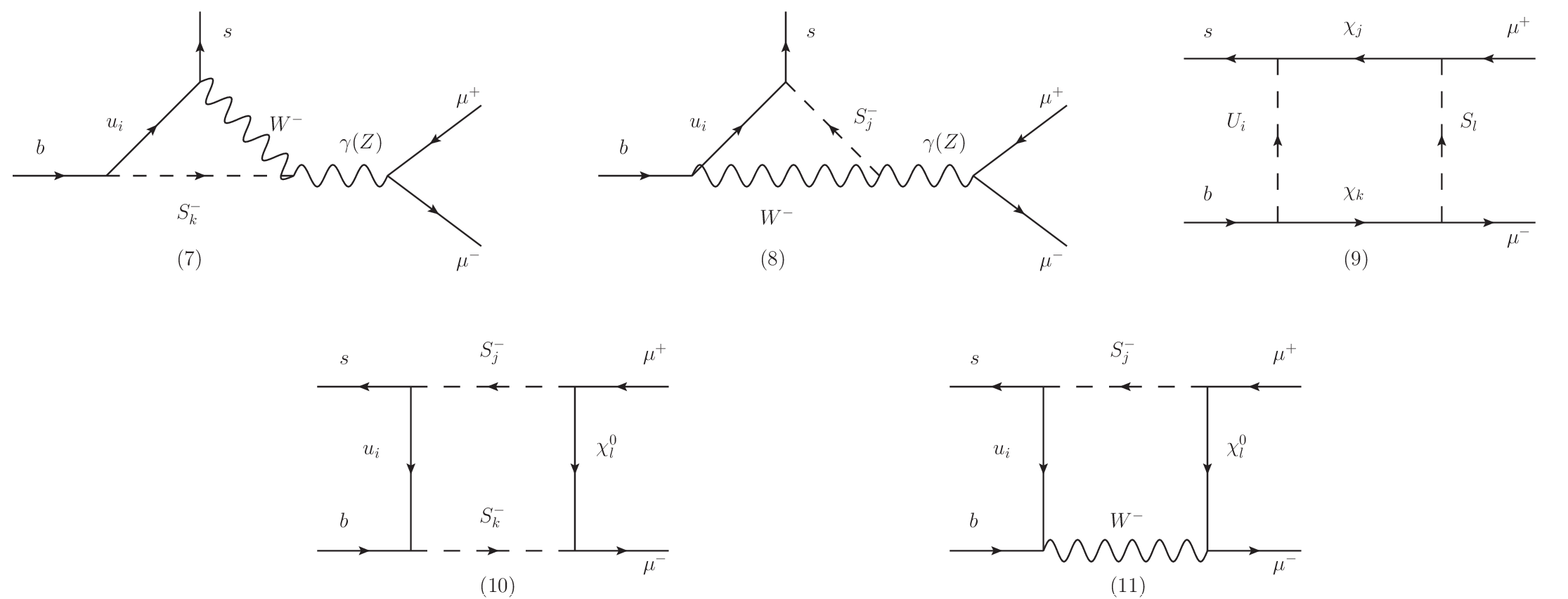}%
	\vspace{0cm}
	\caption[]{The Feynman diagrams contributing to $B_s^0\rightarrow\mu^+\mu^-$ from exotic fields in the $\mu \nu$SSM, compared with the SM.} \label{Bmumu}
\end{figure}

Since the contribution of ${\cal H}_{eff}^{(u)}$ is doubly Cabibbo-suppressed compared to the contribution of ${\cal H}_{eff}^{(t)}$, we can ignore it in the subsequent calculations. The new effective Hamiltonian can be written as:
\begin{eqnarray}
	&&H_{eff}=-\frac{4G_F}{\sqrt{2}}V_{ts}^\ast V_{tb}\Big[C_1\mathcal{O}^c_1+C_2\mathcal{O}_2^c+\sum_{i=3}^6C_i\mathcal{O}_i+\sum_{i=7}^{10}(C_i\mathcal{O}_i+C'_i\mathcal{O}'_i)\nonumber\\
	&&\qquad\;\quad\;+\sum_{i=S,P}(C_i\mathcal{O}_i+C'_i\mathcal{O}'_i)\Big],
\end{eqnarray}
where $\mathcal{O}_i(i=1, 2,...,10, S, P)$ and $\mathcal{O}'_i(i=7, 8,...,10, S, P)$ are defined as follows:
\begin{eqnarray}
	&&{\cal O}_{_1}^c=(\bar{s}_{_L}\gamma_\mu T^ac_{_L})(\bar{c}_{_L}\gamma^\mu T^ab_{_L})\;,\;\;
	{\cal O}_{_2}^c=(\bar{s}_{_L}\gamma_\mu c_{_L})(\bar{c}_{_L}\gamma^\mu b_{_L})\;,
	\nonumber\\
	&&{\cal O}_{_3}=(\bar{s}_{_L}\gamma_\mu b_{_L})\sum\limits_q(\bar{q}\gamma^\mu q)\;,\;\;
	{\cal O}_{_4}=(\bar{s}_{_L}\gamma_\mu T^ab_{_L})\sum\limits_q(\bar{q}\gamma^\mu T^aq)\;,
	\nonumber\\
	&&{\cal O}_{_5}=(\bar{s}_{_L}\gamma_\mu\gamma_\nu\gamma_\rho b_{_L})\sum\limits_q(\bar{q}\gamma^\mu
	\gamma^\nu\gamma^\rho q)\;,\;\;
	{\cal O}_{_6}=(\bar{s}_{_L}\gamma_\mu\gamma_\nu\gamma_\rho T^ab_{_L})\sum\limits_q(\bar{q}\gamma^\mu
	\gamma^\nu\gamma^\rho T^aq)\;,
	\nonumber\\
	&&{\cal O}_{_7}={e\over 16\pi^2}m_{_b}(\bar{s}_{_L}\sigma_{_{\mu\nu}}b_{_R})F^{\mu\nu}\;,\;\;
	{\cal O}_{_7}'={e\over 16\pi^2}m_{_b}(\bar{s}_{_R}\sigma_{_{\mu\nu}}b_{_L})F^{\mu\nu}\;,\;\;
	\nonumber\\
	&&{\cal O}_{_8}={g_{_s}\over 16\pi^2}m_{_b}(\bar{s}_{_L}\sigma_{_{\mu\nu}}T^ab_{_R})G^{a,\mu\nu}\;,\;\;
	{\cal O}_{_8}'={g_{_s}\over 16\pi^2}m_{_b}(\bar{s}_{_R}\sigma_{_{\mu\nu}}T^ab_{_L})G^{a,\mu\nu}\;,\;\;
	\nonumber\\
	&&{\cal O}_{_9}={e^2\over g_{_s}^2}(\bar{s}_{_L}\gamma_\mu b_{_L})\bar{l}\gamma^\mu l\;,\;\;
	{\cal O}_{_9}'={e^2\over g_{_s}^2}(\bar{s}_{_R}\gamma_\mu b_{_R})\bar{l}\gamma^\mu l\;,\;\;
	\nonumber\\
	&&{\cal O}_{_{10}}={e^2\over g_{_s}^2}(\bar{s}_{_L}\gamma_\mu b_{_L})\bar{l}\gamma^\mu\gamma_5 l\;,\;\;
	{\cal O}_{_{10}}'={e^2\over g_{_s}^2}(\bar{s}_{_R}\gamma_\mu b_{_R})\bar{l}\gamma^\mu\gamma_5 l\;,\;\;
	\nonumber\\
	&&{\cal O}_{_S}={e^2\over16\pi^2}m_{_b}(\bar{s}_{_L}b_{_R})\bar{l}l\;,\;\;
	{\cal O}_{_S}'={e^2\over16\pi^2}m_{_b}(\bar{s}_{_R}b_{_L})\bar{l}l\;,\;\;\nonumber\\
	&&{\cal O}_{_P}={e^2\over16\pi^2}m_{_b}(\bar{s}_{_L}b_{_R})\bar{l}\gamma_5l\;,\;\;
	{\cal O}_{_P}'={e^2\over16\pi^2}m_{_b}(\bar{s}_{_R}b_{_L})\bar{l}\gamma_5l\;.
	\label{operators}
\end{eqnarray}
Here, $g_s$ represents the strong coupling, $F^{\mu\nu}$ refers to the electromagnetic field strength tensors, $G^{\mu\nu}$ denotes the gluon field strength tensors, and $T^a\,(a=1,...,8)$ are the generators of $SU(3)$.

The primary Feynman diagrams that contribute to the process $B_s^0\rightarrow\mu^+\mu^-$ in the $\mu\nu{\rm SSM}$ are depicted in Fig.~\ref{Bmumu}, where $S_{l}$ ($l=1,\ldots,8$) denote neutral scalars, $\chi_{l}^0$ ($l=1,\ldots,10$) denote neutral fermions, $S_{i}^-$ ($i=2,\ldots,8$) denote charged scalars, $U_{i}$ ($i=1,\ldots,6$) denote up-type squarks, $u_{i}$ ($i=1,2,3$) denote three generation of up-type quarks and $\chi_{i}$ ($i=1,\ldots,5$) denote charged fermions. At the electroweak energy scale \( \mu_{\text{EW}} \), the corresponding Wilson coefficients can be denoted as
\begin{eqnarray}
	&&C_{_{S,NP}}(\mu_{_{\rm EW}})=\frac{\sqrt{2}s_{_W}c_{_W}}{4m_be^3V_{ts}^*V_{tb}}\Big[C_{_{S,NP}}^{(1)}(\mu_{_{\rm EW}})+C_{_{S,NP}}^{(2)}(\mu_{_{\rm EW}})+C_{_{S,NP}}^{(3)}(\mu_{_{\rm EW}})+C_{_{S,NP}}^{(4)}(\mu_{_{\rm EW}})\nonumber\\
	&&\qquad\;\qquad\;\qquad\;+C_{_{S,NP}}^{(6)}(\mu_{_{\rm EW}})+C_{_{S,NP}}^{(9)}(\mu_{_{\rm EW}})+C_{_{S,NP}}^{(11)}(\mu_{_{\rm EW}})\Big],\nonumber\\
	&&C_{_{S,NP}}^\prime(\mu_{_{\rm EW}})=C_{_{S,NP}}(\mu_{_{\rm EW}})(L\leftrightarrow R),\nonumber\\
	&&C_{_{P,NP}}(\mu_{_{\rm EW}})=\frac{\sqrt{2}s_{_W}c_{_W}}{4m_be^3V_{ts}^*V_{tb}}\Big[C_{_{P,NP}}^{(1)}(\mu_{_{\rm EW}})+C_{_{P,NP}}^{(2)}(\mu_{_{\rm EW}})+C_{_{P,NP}}^{(3)}(\mu_{_{\rm EW}})+C_{_{P,NP}}^{(4)}(\mu_{_{\rm EW}})\nonumber\\
	&&\qquad\;\qquad\;\qquad\;+C_{_{P,NP}}^{(6)}(\mu_{_{\rm EW}})+C_{_{P,NP}}^{(9)}(\mu_{_{\rm EW}})+C_{_{P,NP}}^{(11)}(\mu_{_{\rm EW}})\Big],\nonumber\\
	&&C_{_{P,NP}}^\prime(\mu_{_{\rm EW}})=-C_{_{P,NP}}(\mu_{_{\rm EW}})(L\leftrightarrow R),\nonumber\\
	&&C_{_{9,NP}}(\mu_{_{\rm EW}})=\frac{\sqrt{2}s_{_W}c_{_W}g_{_s}^2}{64\pi^2e^3V_{ts}^*V_{tb}}\Big[C_{_{9,NP}}^{(5)}(\mu_{_{\rm EW}})+C_{_{9,NP}}^{(6)}(\mu_{_{\rm EW}})+C_{_{9,NP}}^{(7)}(\mu_{_{\rm EW}})+C_{_{9,NP}}^{(8)}(\mu_{_{\rm EW}})\nonumber\\
	&&\qquad\;\qquad\;\qquad\;+C_{_{9,NP}}^{(9)}(\mu_{_{\rm EW}})+C_{_{9,NP}}^{(10)}(\mu_{_{\rm EW}})\Big]\;,\nonumber\\
	&&C_{_{9,NP}}^\prime(\mu_{_{\rm EW}})=C_{_{9,NP}}(\mu_{_{\rm EW}})(L\leftrightarrow R),\nonumber\\
	&&C_{_{10,NP}}(\mu_{_{\rm EW}})=\frac{\sqrt{2}s_{_W}c_{_W}g_{_s}^2}{64\pi^2e^3V_{ts}^*V_{tb}}\Big[C_{_{10,NP}}^{(5)}(\mu_{_{\rm EW}})+C_{_{10,NP}}^{(6)}(\mu_{_{\rm EW}})+C_{_{10,NP}}^{(7)}(\mu_{_{\rm EW}})+C_{_{10,NP}}^{(8)}(\mu_{_{\rm EW}})\nonumber\\
	&&\qquad\;\qquad\;\qquad\;+C_{_{10,NP}}^{(9)}(\mu_{_{\rm EW}})+C_{_{10,NP}}^{(10)}(\mu_{_{\rm EW}})\Big]\;,\nonumber\\
	&&C_{_{10,NP}}^\prime(\mu_{_{\rm EW}})=-C_{_{10,NP}}(\mu_{_{\rm EW}})(L\leftrightarrow R).
	\label{Wilson-Coefficients}
\end{eqnarray}
Here, the superscripts (1, ..., 11) correspond to the new physics corrections in Fig.~\ref{Bmumu}, and the specific expressions for these Wilson coefficients are detailed in Appendix~\ref{wilsonbmumu}. The Wilson coefficients at the hadronic energy scale, ranging from the SM to next-to-next-to-logarithmic (NNLL) accuracy, are presented in Table I~\cite{Altmannshofer:2008dz}.
\begin{table}
	\begin{tabular}{|c|c|c|c|}
		\hline
		\hline
		$C_{_7}^{eff,SM}$    & $C_{_8}^{eff,SM}$    & $C_{_9}^{eff,SM}$ & $C_{_{10}}^{eff,SM}$\\
		\hline
		$-0.304$   & $-0.167$  & $4.211$ & $-4.103$\\
		\hline
		\hline
	\end{tabular}
	\caption{At hadronic scale $\mu\sim m_{_b}$, Wilson coefficients from the SM to NNLL accuracy. \label{tab1}}
\end{table}
Furthermore, the Wilson coefficients in Eqs.(\ref{Wilson-Coefficients}) are computed at the matching scale $\mu_{EW}$ and subsequently evolved down to the hadronic scale \( \mu \sim m_b \) through the renormalization group equations:
\begin{eqnarray}
	&&\overrightarrow{C}_{_{NP}}(\mu)=\widehat{U}(\mu,\mu_0)\overrightarrow{C}_{_{NP}}(\mu_0)
	\nonumber\\
	&&\overrightarrow{C^\prime}_{_{NP}}(\mu)=\widehat{U^\prime}(\mu,\mu_0)
	\overrightarrow{C^\prime}_{_{NP}}(\mu_0)\;,
	\label{evaluation1}
\end{eqnarray}
where
\begin{eqnarray}
	&&\overrightarrow{C}_{_{NP}}^{T}=\Big(C_{_{1,NP}}^{eff},\;\cdots,\;C_{_{6,NP}}^{eff},
	C_{_{7,NP}}^{eff},\;C_{_{8,NP}}^{eff},\;C_{_{9,NP}}^{eff},\;
	C_{_{10,NP}}^{eff}\Big)
	\;,\nonumber\\
	&&\overrightarrow{C}_{_{NP}}^{\prime,\;T}=\Big(C_{_{7,NP}}^{\prime,\;eff},\;
	C_{_{8,NP}}^{\prime,\;eff},\;C_{_{9,NP}}^{\prime,\;eff},\;
	C_{_{10,NP}}^{\prime,\;eff}\Big)\;.
	\label{evaluation2}
\end{eqnarray}
According to Ref. \cite{Gambino1}, the definitions of $C_i^{\text{eff}}$ are as follows:	
\begin{eqnarray}
	C_i^{\text{eff}}(\mu) =
	\begin{cases}
		C_i(\mu), & \text{for } i = 1\text{--}6, \\[10pt]
		\frac{4\pi}{\alpha_s} C_i(\mu) + \sum_{j=1}^{6} y_j^{(i)} C_j(\mu), & \text{for } i = 7\text{--}8, \\[10pt]
		\frac{4\pi}{\alpha_s} C_i(\mu) , & \text{for $i = 9$--$10$}.
	\end{cases}
\end{eqnarray}
where $y^{(7)} = (0,0,-\frac{1}{3}, -\frac{4}{9}, -\frac{20}{3}, -\frac{80}{9})$ and
$y^{(8)} = (0, 0, 1, -\frac{1}{6}, 20,-\frac{10}{3})$. Note that the definition of $C_9^{\text{eff}}$ in this work differs slightly from that in Ref.~\cite{Altmannshofer:2008dz} which includes factorisable contributions from the quark loops: \(Y(q^2)\).
Correspondingly, the evolving matrices are approximated as
\begin{eqnarray}
	&&\widehat{U}(\mu,\mu_0)\simeq1-\Big[{1\over2\beta_0}\ln{\alpha_{_s}(\mu)\over
		\alpha_{_s}(\mu_0)}\Big]\widehat{\gamma}^{{\rm eff} (0),\;T}
	\;,\nonumber\\
	&&\widehat{U^\prime}(\mu,\mu_0)\simeq1-\Big[{1\over2\beta_0}\ln{\alpha_{_s}(\mu)\over
		\alpha_{_s}(\mu_0)}\Big]\widehat{\gamma^\prime}^{{\rm eff} (0),\;T}\;.
	\label{evaluation3}
\end{eqnarray}
By utilizing the Eq.~(30) from Ref.~\cite{Gambino1}, we can calculate the corresponding anomalous dimension matrices
\begin{eqnarray}
	&&\widehat{\gamma}^{{\rm eff} (0)}=\left(\begin{array}{cccccccccc}
		-4&{8\over3}&0&-{2\over9}&0&0&-{208\over243}&{173\over162}&-{2272\over729}&0\\
		12&0&0&{4\over3}&0&0&{416\over81}&{70\over27}&{1952\over243}&0\\
		0&0&0&-{52\over3}&0&2&-{176\over81}&{14\over27}&-{6752\over243}&0\\
		0&0&-{40\over9}&-{100\over9}&{4\over9}&{5\over6}&-{152\over243}&-{587\over162}&-{2192\over729}&0\\
		0&0&0&-{256\over3}&0&20&-{6272\over81}&{6596\over27}&-{84032\over243}&0\\
		0&0&-{256\over9}&{56\over9}&{40\over9}&-{2\over3}&{4624\over243}&{4772\over81}&-{37856\over729}&0\\
		0&0&0&0&0&0&{32\over3}&0&0&0\\
		0&0&0&0&0&0&-{32\over9}&{28\over3}&0&0\\
		0&0&0&0&0&0&0&0&0&0\\
		0&0&0&0&0&0&0&0&0&0\\
	\end{array}\right)
	\;,
\end{eqnarray}
\begin{eqnarray}
	&&\widehat{\gamma^\prime}^{{\rm eff} (0)}=\left(\begin{array}{cccc}
		{32\over3}&0&0&0\\
		-{32\over9}&{28\over3}&0&0\\
		0&0&0&0\\0&0&0&0\\
	\end{array}\right)\;.
	\label{ADM1}
\end{eqnarray}

Then, the squared amplitude can be denoted as
\begin{eqnarray}
	&&|\mathcal{M}_s|^2=16G_F^2|V_{tb}V_{ts}^*|^2M_{B_s^0}^2\Big[|F_S^s|^2+|F_P^s+2m_{\mu}F_A^s|^2\Big],
\end{eqnarray}
and
\begin{eqnarray}
	&&F_S^s=\frac{\alpha_{EW}(\mu_b)}{8\pi}\frac{m_b M_{B_s^0}^2}{m_b+m_s}f_{B_s^0}(C_S-C_S'),\\
	&&F_P^s=\frac{\alpha_{EW}(\mu_b)}{8\pi}\frac{m_b M_{B_s^0}^2}{m_b+m_s}f_{B_s^0}(C_P-C_P'),\\
	&&F_A^s=\frac{\alpha_{EW}(\mu_b)}{8\pi}f_{B_s^0}\Big[C_{10}^{eff}(\mu_b)-C_{10}^{\prime eff}(\mu_b)\Big].
\end{eqnarray}
Here, the decay constant is denoted by \( f_{B^0_s} = 230.3\,(1.3) \text{MeV} \)~\cite{FlavourLatticeAveragingGroupFLAG:2021npn,Bazavov:2017lyh,ETM:2016nbo,Dowdall:2013tga,Hughes:2017spc}, and the mass of the neutral meson \( B_s^0 \) is represented by \( M_{B^0_s} =5366.93\,(\pm0.10) \text{MeV} \)~\cite{ParticleDataGroup:2024cfk}.

Ultimately, the branching ratio of \( B_s^0 \rightarrow \mu^+\mu^- \) can be expressed as:
\begin{eqnarray}
&&Br(B_s^0\rightarrow\mu^+\mu^-)=\frac{\tau_{B_s^0}}{16\pi}\frac{|\mathcal{M}_s|^2}{M_{B_s^0}}\sqrt{1-\frac{4m_{\mu}^2}{M_{B_s^0}^2}},
\end{eqnarray}
where \(\tau_{B_s^0} = 1.527\,(\pm0.011)\) ps~\cite{HFLAV:2022esi} denotes the lifetime of \( B_s^0 \).

\section{Numerical analysis\label{sec4}}
\indent\indent
In this section, we provide the numerical results of the branching ratios for rare \( B \) meson decays \( B_s^0 \rightarrow \mu^+\mu^- \) and \( \bar B \rightarrow X_s\gamma \). We analyzed how individual parameters affect the branching ratios of these two processes. To better understand the impact of these parameters on the branching ratios, we need to first make a reasonable selection of other parameters.

The relevant SM input parameters are presented in Table ~\ref{SMnuminput}. All other parameters in SM remain unchanged compared to those listed in Table I of Ref.~\cite{Bobeth:2013uxa}, as their modification would have negligible impact on \(Br(B_s^0 \rightarrow \mu^+\mu^-) \) either because they are already measured with high precision or because their influence on \(Br(B_s^0 \rightarrow \mu^+\mu^-) \) is minimal.

\begin{table}[t]
	\begin{center}
		\begin{tabular}{llcl}
			\textbf{Parameter}
			& \textbf{Value}
			& \textbf{Unit}
			& \textbf{Ref.}
			\\
			\hline
			$\tau_{B^0_s}$
			& $1.527\,(\pm0.011)$
			& ps
			& \cite{HFLAV:2022esi}
			\\
			$\alpha_s(m_Z)$
			& $0.1180\, (9)$
			& --
			& \cite{ParticleDataGroup:2024cfk}	
			\\
			$\alpha_{em}(m_Z)$
			& $1/127.944\, (14)$
			& --
			& \cite{ParticleDataGroup:2024cfk}
			\\
			$M_{B^0_s}$
			& $5366.93\,(\pm0.10)$
			& MeV
			& \cite{ParticleDataGroup:2024cfk}
			\\
			$f_{B^0_s}$
			& $230.3\,(1.3)$
			& MeV
			& \cite{FlavourLatticeAveragingGroupFLAG:2021npn,Bazavov:2017lyh,ETM:2016nbo,Dowdall:2013tga,Hughes:2017spc}
			\\
			$|V_{cb}|\times10^3$
			& $41.97\,(48)$
			& --
			& \cite{Finauri:2023kte}
			\\
			$|V_{tb}^\star V_{ts}^{}/V_{cb}^{}|$
			& $0.9820\,(4)$
			& --
			& \cite{Charles:2004jd}
			\\
			\hline
		\end{tabular}
		\caption{ Numerical values of the updated input parameters in the SM. \label{SMnuminput} }
	\end{center}
\end{table}

In the SUSY extensions of the SM, there exist numerous free parameters. Given the structure of the soft SUSY-breaking terms, the free parameters in the $\mu\nu{\rm SSM}$ are:
\begin{eqnarray}
&&\lambda_i,\,\,
\kappa_{ijk},\,\,
Y_{\nu ij},\,\,
m_{H_d}^2,\,\,
m_{H_u}^2,\,\,
m_{\tilde \nu_{ij}^c}^2,\,\,
m_{{{\tilde L}_{ij}}}^2,\,\,
({A_\lambda }\lambda )_i,\,\,
({A_\kappa }\kappa )_{ijk},\,\,
(A_\nu Y_\nu)_{ij}, \nonumber\\
&&M_1, M_2, M_3,\,\,
m_{\tilde Q_{ij}}^2,\,\,
m_{\tilde u_{ij}^c}^2,\,\, 	
m_{\tilde d_{ij}^c}^2,\,\,
m_{\tilde e_{ij}^c}^2,\,\,
(A_u Y_u)_{ij},\,\,
(A_d Y_d)_{ij},\,\,
({A_e}{Y_e})_{ij}.
\label{all parameters}
\end{eqnarray}

To streamline numerical results, we apply the minimal flavor violation (MFV) assumption to certain parameters in the $\mu\nu{\rm SSM}$. This assumption includes:
\begin{eqnarray}
	&&\lambda _i = \lambda , \qquad {\kappa _{ijk}} = \kappa {\delta _{ij}}{\delta _{jk}}, \quad  {({A_\kappa }\kappa )_{ijk}} ={A_\kappa }\kappa {\delta _{ij}}{\delta _{jk}}, \nonumber\\
	&&{({A_\lambda }\lambda )}_i= {A_\lambda }\lambda,\quad {Y_{{\nu _{ij}}}} = {Y_{{\nu _i}}}{\delta _{ij}},\qquad {Y_{{e_{ij}}}} = {Y_{{e_i}}}{\delta _{ij}},
	\nonumber\\
	&&\upsilon_{\nu_i^c}=\upsilon_{\nu^c},\quad(A_\nu Y_\nu)_{ij}={a_{{\nu_i}}}{\delta _{ij}},\quad{({A_e}{Y_e})_{ij}} = {A_e}{Y_{{e_i}}}{\delta _{ij}},\nonumber\\
	&&m_{{{\tilde L}_{ij}}}^2 = m_{\tilde L}^2{\delta _{ij}},\quad
	m_{\tilde \nu_{ij}^c}^2 = m_{{{\tilde \nu_i}^c}}^2{\delta _{ij}},\quad
	m_{\tilde e_{ij}^c}^2 = m_{{{\tilde e}^c}}^2{\delta _{ij}},\nonumber\\
	&&m_{\tilde Q_{ij}}^2 = m_{{{\tilde Q}_i}}^2{\delta _{ij}}, \quad
	m_{\tilde u_{ij}^c}^2 = m_{{{{\tilde u}_i}^c}}^2{\delta _{ij}}, \quad
	m_{\tilde d_{ij}^c}^2 = m_{{{{\tilde d}_i}^c}}^2{\delta _{ij}}.
	\label{assumption}
\end{eqnarray}
For the relevant parameters for quarks, one can have
\begin{eqnarray}
	&&(A_u Y_u)_{ij}={A_{u_i}}{Y_{{u_{ij}}}},\quad {Y_{{u _{ij}}}} = {Y_{{u _i}}}{V_{L_{ij}}^u},\nonumber\\
	&&(A_d Y_d)_{ij}={A_{d_i}}{Y_{{d_{ij}}}},\quad {Y_{{d_{ij}}}} = {Y_{{d_i}}}{V_{L_{ij}}^d},
\end{eqnarray}
where $V=V_L^u V_L^{d\dag}$ denotes the CKM matrix~\cite{CKM1,CKM2}. Constrained by the masses of quarks and leptons, approximate relations can be defined: \( Y_{u_i} \approx \frac{m_{u_i}}{\upsilon_u} \), \( Y_{d_i} \approx \frac{m_{d_i}}{\upsilon_d} \), \( Y_{e_i} = \frac{m_{l_i}}{\upsilon_d} \), where \( m_{u_i} \), \( m_{d_i} \), and \( m_{l_i} \) denote the masses of up-quarks, down-quarks, and charged leptons, respectively. The specific values are adopted from PDG.~\cite{ParticleDataGroup:2024cfk}.
Additionally, the aforementioned soft masses $m_{H_d}^2, m_{H_u}^2, m_{\tilde \nu_{ij}^c}^2$ and $m_{{{\tilde L}_{ij}}}^2$ can be substituted with the VEVs in Eq.~(\ref{VEVs}). By utilizing \( \tan \beta \equiv \frac{v_u}{v_d} \) and the SM Higgs VEV, \( v^2 = v_d^2 + v_u^2 + \sum_i \upsilon_{\nu_i}^2 = \frac{2m_Z^2}{(g_1^2 + g_2^2)} \approx (174 \text{ GeV})^2 \) with the electroweak gauge couplings estimated at the \( m_Z \) scale by \( e = g_1 \cos \theta_W = g_2 \sin \theta_W \), one can derive the SUSY Higgs VEVs, \( v_d \) and \( v_u \). Given that \( \upsilon_{\nu_i} \ll v_d, v_u \), it follows that \( v_d \approx \frac{v}{\sqrt{\tan^2 \beta + 1}} \).

In our prior investigation~\cite{neu-mass6}, we extensively examined how the neutrino oscillation data restrict neutrino Yukawa couplings \( Y_{\nu_i} \sim \mathcal{O}(10^{-7}) \) and left-handed sneutrino VEVs \( \upsilon_{\nu_i} \sim \mathcal{O}(10^{-4} \, \text{GeV}) \) within the $\mu\nu$SSM through the TeV-scale seesaw mechanism. As the tiny neutrino masses have minimal impact on \( Br(\bar B \rightarrow X_s\gamma) \) and \(Br(B_s^0 \rightarrow \mu^+\mu^-) \), we can approximate \( Y_{\nu_i}= 0 \) and \(\upsilon_{\nu_i}=0\). For the Majorana masses of the gauginos, we will imply the approximate GUT relation $M_1=\frac{\alpha_1^2}{\alpha_2^2}M_2\approx 0.5 M_2$ and $M_3=\frac{\alpha_3^2}{\alpha_2^2}M_2\approx 2.7 M_2$. The gluino mass, $m_{{\tilde g}}\approx M_3$, is larger than about $1.2$ TeV from the ATLAS and CMS experimental data~\cite{ATLAS-sg1,ATLAS-sg2,CMS-sg1,CMS-sg2}. So, we conservatively choose $M_2=1\;{\rm TeV}$. Finally, the free parameters in Eq.~(\ref{all parameters}) have been replaced with the following:
\begin{eqnarray}
	&&\lambda,\,\,
	\kappa,\,\,
	\tan \beta,\,\,
	\upsilon_{\nu^c},\,\,
	{A_\lambda },\,\,
	{A_\kappa },\nonumber\\
	&&m_{\tilde Q_{i}}^2,\,\,
	A_{d_i},\,\,
	m_{\tilde d_{i}^c}^2,\,\,
	m_{\tilde u_{i}^c}^2,\,\,
	m_{\tilde e_{i}^c}^2,\,\,
	A_{u_i},\,\,
	A_{e_i}.
	\label{all parameters2}
\end{eqnarray}

Among the parameters in Eq.~(\ref{all parameters2}), $ \kappa, \tan \beta, \upsilon_{\nu^c}, A_\lambda$, and $A_{u_3}$ have a significant impact on the results. The remaining parameters of the model are fixed as shown in Table ~\ref{fixed-parameters}.
\begin{table}
	\begin{tabular}{|cccccc|}
		\hline
		\multicolumn{3}{|c|}{$m_{\tilde Q_{1,2}}=m_{\tilde u_{1,2}^c}=m_{\tilde d_{1,2}^c}=3$ TeV} \\
		\multicolumn{3}{|c|}{\( Y_{\nu_i}= 0 \), \(\upsilon_{\nu_i}=0\)}\\
		\multicolumn{3}{|c|}{$m_{{\tilde Q}_3}=m_{{\tilde u}^c_3}=m_{{\tilde d}^c_3}=2$ TeV}\\
		\multicolumn{3}{|c|}{$\lambda = 0.05$, $m_{\tilde e_{1,2,3}^c}=1$ TeV}\\
		\multicolumn{3}{|c|}{$A_{u_{1,2}}=A_{e_{1,2,3}}=A_{d_{1,2,3}}=1$ TeV}\\
		\multicolumn{3}{|c|}{$M_1= 0.5 M_2$, $M_3=2.7 M_2$, $M_2 = 1$ TeV}\\
		\multicolumn{3}{|c|}{$A_{\kappa} = -300$ GeV} \\
		\hline
	\end{tabular}
	\caption{This table summarizes the model parameters and values.}
	\label{fixed-parameters}
\end{table}
For squarks, the first two generations of squarks are strongly constrained by direct searches at the Large Hadron Collider (LHC)~\cite{ATLAS.PRD,CMS.JHEP}. Therefore, we take $m_{{\tilde Q}_{1,2}}=m_{{\tilde u}^c_{1,2}}=m_{{\tilde d}^c_{1,2}}=3\;{\rm TeV}$. The third generation squark masses are not as strictly constrained by the LHC as the first two generations. Additionally, the smaller the mass of the third-generation squarks, the more pronounced the effect of the parameter $A_{u_{3}}$ on \(Br(B_s^0 \rightarrow \mu^+\mu^-) \). To clearly investigate the impact of $A_{u_{3}}$ on \(Br(B_s^0 \rightarrow \mu^+\mu^-) \), we have set the mass of the third-generation squarks slightly lower, specifically $m_{{\tilde Q}_3}=m_{{\tilde u}^c_3}=m_{{\tilde d}^c_3}=2\;{\rm TeV}$.

The variations in \(m_{\tilde e_{i}^c}^2\), \(A_{e_{i}}\), \(A_{d_{i}}\), and \(A_{u_{1,2}}\) have a negligible impact on the calculation results for \(Br(B_s^0 \rightarrow \mu^+\mu^-) \) and \( Br(\bar B \rightarrow X_s\gamma) \). Based on the parameter space analysis outlined in Ref.~\cite{ref3}, we can select reasonable values for certain parameters: \( \lambda = 0.05 \), \(A_{d_{1,2,3}} = A_{e_{1,2,3}} = A_{u_{1,2}} = 1 \, \text{TeV}\), and \(m_{\tilde e_{i}^c} = 1 \, \text{TeV}\) for simplicity in subsequent numerical computations. The parameter $A_{u_3}=A_t$ is a key factor influencing the subsequent numerical calculations.

In the $\mu\nu{\rm SSM}$, the sneutrino sector may exhibit tachyonic behavior. The squared masses of these tachyons are negative. Therefore, it is necessary to analyze the masses of the sneutrinos. The masses of the left-handed sneutrinos are primarily determined by \(m_{\tilde{L}}\), while the three right-handed sneutrinos are essentially degenerate. The mass squared for the CP-even and CP-odd right-handed sneutrinos can be approximated as follows~\cite{ref-zhang1}:
\begin{eqnarray}
	&&m_{S_{5+i}}^2\approx (A_\kappa+4\kappa\upsilon_{\nu^c})\kappa\upsilon_{\nu^c} +A_\lambda \lambda \upsilon_d \upsilon_u/\upsilon_{\nu^c}-2\lambda^2(\upsilon_d^2+\upsilon_u^2),\\
	&&m_{P_{5+i}}^2\approx -3A_\kappa \kappa\upsilon_{\nu^c} +(A_\lambda/\upsilon_{\nu^c}+4\kappa)\lambda \upsilon_d \upsilon_u-2\lambda^2(\upsilon_d^2+\upsilon_u^2).
\end{eqnarray}
In these expressions, the primary contribution to the mass squared comes from the first term when \( \kappa \) is large, in the limit of \( \upsilon_{\nu^{c}} \gg \upsilon_{u,d} \). Hence, we can use the approximate relation
\begin{eqnarray}
	-4\kappa\upsilon_{\nu^c}\lesssim A_\kappa \lesssim 0,
	\label{tachyon}
\end{eqnarray}
to prevent the occurrence of tachyons. While $A_\kappa$ has minimal influence on the computation results, both $\kappa$ and $\upsilon_{\nu^c}$ are crucial parameters affecting the calculations. In the subsequent analysis, $\kappa$ varies between 0.1 and 1, while $\upsilon_{\nu^c}$ ranges from 1.5TeV to 2.5TeV. According to Eq.~(\ref{tachyon}), we can conservatively set \(A_{\kappa} = -300 \, \text{GeV}\) to avoid the emergence of tachyons.

In the limit of \( \upsilon_{\nu^c} \gg \upsilon_{u,d} \) \cite{ref-limit-MH}, the squared mass of the charged Higgs \( M_{H^\pm}^2 \) in the $\mu \nu$SSM can be expressed as:
\begin{eqnarray}
	M_{H^\pm}^2\simeq M_A^2+(1-\frac{6s_{_W}^2\lambda^2}{e^2})m_W^2,
	\label{mch1}
\end{eqnarray}
with the squared mass of the neutral pseudoscalar:
\begin{eqnarray}
	M_A^2\simeq \frac{6\lambda\upsilon_{\nu^c}(A_\lambda+\kappa\upsilon_{\nu^c})}{\sin2\beta}.
	\label{mch2}
\end{eqnarray}

\begin{figure}
	\setlength{\unitlength}{1mm}
	\centering
	\includegraphics[width=3.1in]{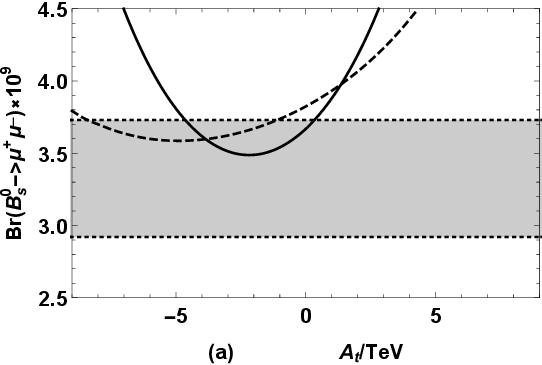}%
	\vspace{0.5cm}
	\includegraphics[width=3.1in]{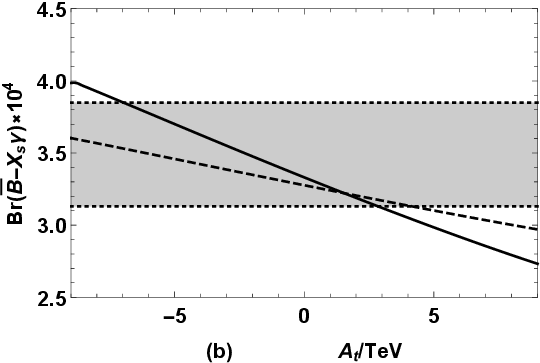}%
	\vspace{0cm}
	\caption[]{When \(\kappa = 0.5\), \(\upsilon_{\nu^c} = 2 \, \text{TeV}\), and \(A_\lambda = 0.5 \, \text{TeV}\), (a) \(Br(B_s^0 \rightarrow \mu^+\mu^-)\) and (b) \(Br(\bar B \rightarrow X_s \gamma)\) vary with \(A_t\) for \(\tan\beta = 30 \, (\text{solid line})\) and \(\tan\beta = 15 \, (\text{dashed line})\).}
	\label{figAt}
\end{figure}

In the scenarios of the MSSM and NMSSM~\cite{Domingo:2007dx}, the additional physics contributions to \( Br(B_s^0 \to \mu^+ \mu^-) \) and \( Br(\bar{B} \to X_s \gamma) \) primarily depend on \( A_t \), \( \tan \beta \), and the charged Higgs mass \( M_{H^\pm} \). To demonstrate the impacts of \( A_t \) and \( \tan \beta \) on the theoretical evaluations of \(Br(B_s^0 \rightarrow \mu^+\mu^-) \) and \( Br(\bar B \rightarrow X_s\gamma) \) in the $\mu\nu{\rm SSM}$, we temporarily fix the parameters $\kappa$, $\upsilon_{\nu^c}$, and $A_\lambda$. Specifically, we set $\kappa = 0.5$, $\upsilon_{\nu^c}=2\;{\rm TeV}$, and $A_\lambda=0.5\;{\rm TeV}$. The plots in Fig.~\ref{figAt} show \( Br(B_s^0 \to \mu^+ \mu^-) \) and \( Br(\bar{B} \to X_s \gamma) \) as functions of \( A_t \) for \( \tan \beta = 30 \) (solid line) and \( \tan \beta = 15 \) (dashed line).

Additionally, since our calculation method essentially combines the new contributions from SUSY particles with the existing SM contributions to obtain the final result, we can roughly consider the theoretical uncertainty in the SM as the minimum uncertainty for the $\mu\nu{\rm SSM}$. We treat the theoretical uncertainty in the SM from Eq.~(\ref{SM bsr}, \ref{SM Buu}) as the minimum uncertainty in the $\mu\nu{\rm SSM}$, which is then combined with the experimental \(1\sigma\) uncertainty, represented by the gray band in the figure. The central value of the gray band is the experimental central value from Eq.~(\ref{experimental data}). The range of the gray band in Fig.~\ref{figAt}(a) is \((3.34\pm0.27\pm0.12)\times 10^{-9} =(3.34\pm0.39)\times 10^{-9}\), while the range of the gray band in Fig.~\ref{figAt}(b) is \((3.49\pm0.19\pm0.17)\times 10^{-4} =(3.49\pm0.36)\times 10^{-4}\).

From Fig.~\ref{figAt}, we can see that both $B_s^0\rightarrow \mu^+\mu^-$ and $\bar B\rightarrow X_s\gamma$ constrain the parameter space of the $\mu\nu{\rm SSM}$. From Fig.~\ref{figAt}(a), it is evident that as \( A_t \) increases, \(Br(B_s^0 \to \mu^+ \mu^-) \) will first decrease before increasing. The experimental data on \(Br(B_s^0 \to \mu^+ \mu^-) \) favors \( A_t \) within specific ranges: \( -8.5 \, \text{TeV} \lesssim A_t \lesssim -1.1 \, \text{TeV}\) for \( \tan \beta = 15 \) and \( -4.5 \, \text{TeV} \lesssim A_t \lesssim 0.5 \, \text{TeV}\) for \( \tan \beta = 30 \). From Fig.~\ref{figAt}(b), it is evident that \( Br(\bar B \to X_s \gamma) \) decreases with the escalation of \( A_t \). As the value of \( \tan \beta\) increases, the curve becomes steeper, consequently narrowing the allowed range of \( A_t \). When \( \tan \beta \) is set to 15, the allowable range for \( A_t \) is \( A_t \lesssim 4 \, \text{TeV} \), and when \( \tan \beta \) is set to 30, the allowable range for \( A_t \) is \( -7 \, \text{TeV} \lesssim A_t \lesssim 2.6 \, \text{TeV} \). The influence of \( A_t \) on the calculation results mainly comes from its ability to adjust the coupling strengths between the stop quark and other particles in the Feynman diagrams.

Considering the constraint from the SM-like Higgs mass, in the upcoming parameter analysis, we set \( A_t \)=$-3.6$ TeV with \( \tan \beta = 15 \) to ensure the SM-like Higgs mass around $125{\rm GeV}$. In addition, $\upsilon_{\nu^c}$ is a unique parameter in the \( \mu\nu \)SSM. To investigate the impact of $\upsilon_{\nu^c}$ on the results, we will plot curves for two cases: $\upsilon_{\nu^c}$=1.5 TeV represented by a solid line and $\upsilon_{\nu^c}$=2 TeV represented by a dashed line.

\begin{figure}
	\setlength{\unitlength}{1mm}
	\centering
	\includegraphics[width=3.1in]{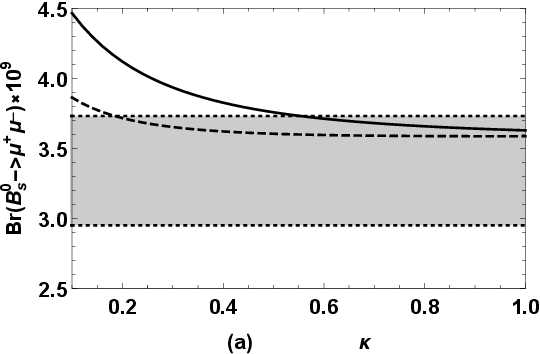}%
	\vspace{0.5cm}
	\includegraphics[width=3.1in]{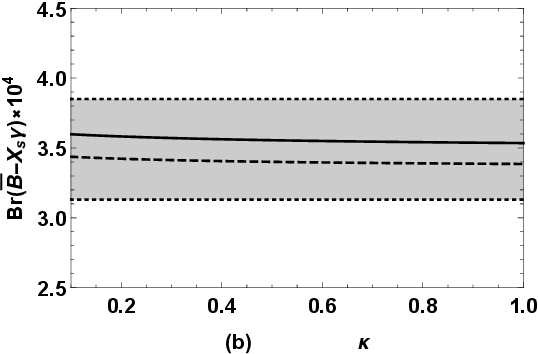}%
	\vspace{0cm}
	\caption[]{When \(\tan\beta = 15, A_t=-3.6\) TeV, and \(A_\lambda = 0.5 \, \text{TeV}\), (a) $Br(B_s^0\rightarrow \mu^+\mu^-)$ and (b) $Br(\bar B\rightarrow X_s\gamma)$ vary with \(\kappa\) for $\upsilon_{\nu^c}=1.5$ TeV (solid line), $\upsilon_{\nu^c}=2$ TeV (dashed line).}
	\label{[Kappa]}
\end{figure}

In the MSSM~\cite{Domingo:2007dx}, the theoretical predictions of \(Br(B_s^0 \rightarrow \mu^+\mu^-) \) and \( Br(\bar B \rightarrow X_s\gamma) \)  will decrease as \( M_{H^\pm} \) increases. From Eqs.(\ref{mch1}, \ref{mch2}), it can be seen that \(\kappa\) can influence \( M_{H^\pm} \) in the $\mu \nu$SSM. Moreover, an increase in \(\kappa\) leads to an increase in \(M_{H^\pm}\). To explore the influences of \(\kappa\) on $Br(B_s^0\rightarrow\mu^+\mu^-)$ and $Br(\bar B\rightarrow X_s\gamma)$, we plot curves in Fig.~\ref{[Kappa]} showing how $Br(\bar B\rightarrow X_s\gamma)$ and $Br(B_s^0\rightarrow\mu^+\mu^-)$ vary with \(\kappa\).
From Fig.~\ref{[Kappa]}(a), it is evident that $Br(B_s^0\rightarrow\mu^+\mu^-)$ gradually decreases as \(\kappa\) increases, eventually stabilizing within the range that is consistent with the experimental data. When  $\upsilon_{\nu^c}$ is set to 2 TeV, the experimental data concerning $Br(B_s^0\rightarrow\mu^+\mu^-)$ favor \(\kappa\) lying within the range \(\kappa \gtrsim 0.18\). When $\upsilon_{\nu^c}$ is set to 1.5 TeV, \(\kappa\) is required to be approximately greater than 0.55.
From Fig.~\ref{[Kappa]}(b), it can be observed that the theoretical predictions for \(Br(\bar B \to X_s \gamma)\) are consistent with the experimental results, and as \(\kappa\) increases, there is a slight decrease in \(Br(\bar B \to X_s \gamma)\). From Fig.~\ref{[Kappa]}, we can also observe that both $Br(B_s^0\rightarrow\mu^+\mu^-)$ and $Br(\bar B\rightarrow X_s\gamma)$ decrease as $\upsilon_{\nu^c}$ increases.
From Eqs.(\ref{mch1}, \ref{mch2}), it can be seen that, similar to \(\kappa\), an increase in $\upsilon_{\nu^c}$ will also lead to an increase in \(M_{H^\pm}\). Therefore, both $\upsilon_{\nu^c}$ and \(\kappa\) can influence the theoretical predictions for $Br(B_s^0\rightarrow\mu^+\mu^-)$ and $Br(\bar B\rightarrow X_s\gamma)$ by affecting \(M_{H^\pm}\). The parameter \(\lambda\) also have effects similar to those of the parameter  $\upsilon_{\nu^c}$.

\begin{figure}
	\setlength{\unitlength}{1mm}
	\centering
	\includegraphics[width=3.1in]{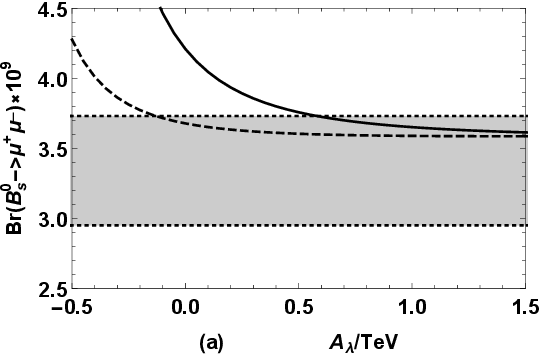}%
	\vspace{0.5cm}
	\includegraphics[width=3.1in]{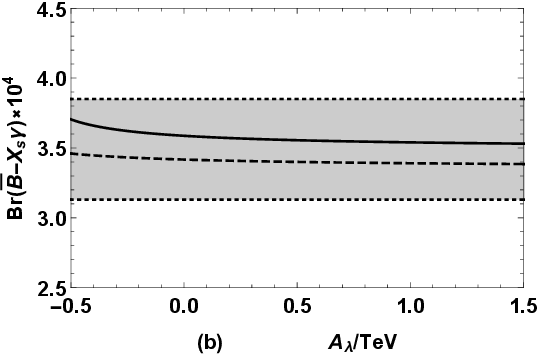}%
	\vspace{0cm}
	\caption[]{When \(\tan\beta = 15, A_t=-3.6\) TeV, and \(\kappa = 0.5\), (a) $Br(B_s^0\rightarrow \mu^+\mu^-)$ and (b) $Br(\bar B\rightarrow X_s\gamma)$ versus \( A_{\lambda} \) for $\upsilon_{\nu^c}=1.5$ TeV (solid line), $\upsilon_{\nu^c}=2$ TeV (dashed line).}
	\label{A[Lambda]}
\end{figure}

From Eqs.(\ref{mch1}, \ref{mch2}), it can be seen that an increase in the parameter \( A_{\lambda} \) will also lead to an increase in \(M_{H^\pm}\).
Thus, we speculate that the parameter \( A_{\lambda} \), like $\upsilon_{\nu^c}$ and \(\kappa\), can also influence the theoretical predictions for $Br(B_s^0\rightarrow\mu^+\mu^-)$ and $Br(\bar B\rightarrow X_s\gamma)$. To investigate this influence, we set \(\kappa\)=0.5, and then plot curves in Fig.~\ref{A[Lambda]} showing how $Br(\bar B\rightarrow X_s\gamma)$ and $Br(B_s^0\rightarrow\mu^+\mu^-)$ vary with \( A_{\lambda}\).
 From Fig.~\ref{A[Lambda]}(a), it is evident that $Br(B_s^0\rightarrow\mu^+\mu^-)$ gradually decreases as \( A_{\lambda}\) increases, eventually stabilizing within the range consistent with the experimental data. The experimental data on $Br(B_s^0\rightarrow\mu^+\mu^-)$ limits that $A_{\lambda}\gtrsim0.58 \, \text{TeV}$ for $\upsilon_{\nu^c}$=1.5 TeV and $A_{\lambda}\gtrsim-0.14 \, \text{TeV}$ for $\upsilon_{\nu^c}$=2 TeV. From Fig.~\ref{A[Lambda]}(b), the theoretical predictions for $Br(\bar B\rightarrow X_s\gamma)$ remain consistent with the experimental results. As \(A_{\lambda}\) increases, $Br(\bar B\rightarrow X_s\gamma)$ will slightly decrease. Similar to the situation in Fig.~\ref{[Kappa]}, an increase of $\upsilon_{\nu^c}$ will result in a decrease in both $Br(B_s^0\rightarrow \mu^+\mu^-)$ and $Br(\bar B\rightarrow X_s\gamma)$, with this effect becoming more pronounced as \( A_{\lambda}\) decreases.

\section{Summary\label{sec6}}
\indent\indent
In this study, we examine the branching ratio of the rare decay $B_s^0\rightarrow\mu^+\mu^-$ within the framework of the \( \mu\nu \)SSM combined with \( \bar{B} \to X_s \gamma \). Similar to the MSSM and NMSSM, the additional physics contributions to $Br(B_s^0\rightarrow\mu^+\mu^-)$ and \(Br(\bar{B} \to X_s \gamma) \) in the \(\mu\nu\)SSM primarily rely on \( M_{H^{\pm}} \), \( \tan \beta \), and \( A_t \). This is due to the suppressed mixings between charginos and charged leptons in the mass matrix of the \( \mu\nu \)SSM, as well as the mixings between charged Higgs bosons and charged sleptons. Additionally, \(\upsilon_{\nu^c}\), \(A_{\lambda}\), and \(\kappa\) in the \(\mu\nu\)SSM can also influence the theoretical predictions of \(Br(B_s^0 \to \mu^+\mu^-)\) and \(Br(\bar{B} \to X_s \gamma)\) by affecting \( M_{H^{\pm}} \). Subject to the constraint of a SM-like Higgs boson with a mass approximately at 125 GeV, the numerical findings indicate that the new physics can align with the experimental data for $B$ meson rare decays and consequently narrow down the parameter space. In addition to \(B_s^0 \rightarrow \mu^+\mu^-\) and \(B \to X_s \gamma\), the \(\mu\nu\)SSM also contributes to \(C_9\). This contribution may lead to other \(b \to s\) observables, such as the "$B$-anomalies", which we will investigate in detail elsewhere.

\begin{acknowledgments}
\indent\indent
The work has been supported by the National Natural Science Foundation of China (NNSFC) with Grants No. 12075074, No. 12235008, Hebei Natural Science Foundation with Grants No. A2022201017, No. A2023201041, and the youth top-notch talent support program of the Hebei Province.
\end{acknowledgments}

\appendix

\section{The Wilson coefficients of the process $B_s^0\rightarrow \mu^+\mu^-$ in the \( \mu\nu \)SSM. \label{wilsonbmumu}}
In our previous work~\cite{Bsmumu:2018}, we have calculated the Wilson coefficients of the process $B_s^0\rightarrow \mu^+\mu^-$ in the B-LSSM. Because the SUSY particles contained in the B-LSSM and the \( \mu\nu \)SSM are different, the Feynman diagrams involving SUSY particles will vary when calculating the amplitude for the same process \(B_s^0 \rightarrow \mu^+\mu^-\). Through the previous work, the Wilson coefficients corresponding to $b\rightarrow s\mu^+\mu^-$ in the \( \mu\nu \)SSM can be written as
\begin{eqnarray}
	&&C_{_{S,NP}}^{(1)}(\mu_{_{\rm EW}})=\sum_{_{U_i,\chi_j,\chi_k,S_l
	}}\frac{C_{\mu^-S_l\mu^+}^L+C_{\mu^-S_l\mu^+}^R}{2(m_b^2-m^2_{S_l})}\Big[C_{U_i\bar s \chi_j}^R C_{\chi_j S_l \chi_k}^L C_{\chi_k s U_i}^R G_2(x_{U_i},x_{\chi_j},x_{\chi_k})\nonumber\\
	&&\qquad\qquad\qquad+m_{\chi_j}m_{\chi_k} C_{U_i\bar s \chi_j}^R C_{\chi_j S_l \chi_k}^R C_{\chi_k s U_i}^R G_1(x_{U_i},x_{\chi_j},x_{\chi_k})\Big]\nonumber\\
	&&\qquad\qquad\qquad+\sum_{_{S^-_i,u_j,u_k,S_l}}\frac{C_{\mu^-S_l\mu^+}^L+C_{\mu^-S_l\mu^+}^R}{2(m_b^2-m^2_{S_l})}\Big[C_{S^-_i\bar s u_j}^R C_{\bar u_j S_l u_k}^L C_{\bar u_k b S^-_i}^R G_2(x_{S^-_i},x_{u_j},x_{u_k})\nonumber\\
	&&\qquad\qquad\qquad+m_{u_j}m_{u_k}C_{S^-_i\bar s u_j}^R C_{\bar u_j S_l u_k}^R C_{\bar u_k b S^-_i}^R G_1(x_{S^-_i},x_{u_j},x_{u_k})\Big],\nonumber\\
	&&C_{_{P,NP}}^{(1)}(\mu_{_{\rm EW}})=\sum_{_{U_i,\chi_j,\chi_k,S_l
	}}\frac{-C_{\mu^-S_l\mu^+}^L+C_{\mu^-S_l\mu^+}^R}{2(m_b^2-m^2_{S_l})}\Big[C_{U_i\bar s \chi_j}^R C_{\chi_j S_l \chi_k}^L C_{\chi_k s U_i}^R G_2(x_{U_i},x_{\chi_j},x_{\chi_k})\nonumber\\
	&&\qquad\qquad\qquad+m_{\chi_j}m_{\chi_k} C_{U_i\bar s \chi_j}^R C_{\chi_j S_l \chi_k}^R C_{\chi_k s U_i}^R G_1(x_{U_i},x_{\chi_j},x_{\chi_k})\Big]\nonumber\\
	&&\qquad\qquad\qquad+\sum_{_{S^-_i,u_j,u_k,S_l}}\frac{-C_{\mu^-S_l\mu^+}^L+C_{\mu^-S_l\mu^+}^R}{2(m_b^2-m^2_{S_l})}\Big[C_{S^-_i\bar s u_j}^R C_{\bar u_j S_l u_k}^L C_{\bar u_k b S^-_i}^R G_2(x_{S^-_i},x_{u_j},x_{u_k})\nonumber\\
	&&\qquad\qquad\qquad+m_{u_j}m_{u_k}C_{S^-_i\bar s u_j}^R C_{\bar u_j S_l u_k}^R C_{\bar u_k b S^-_i}^R G_1(x_{S^-_i},x_{u_j},x_{u_k})\Big],
\end{eqnarray}
\begin{eqnarray}
	&&C_{_{S,NP}}^{(2)}(\mu_{_{\rm EW}})=\sum_{u_i,S^-_j,S^-_k,S_l}\frac{1}{2(m_b^2-m^2_{S_l})}m_{u_i}C_{\bar s u_i S^-_j}^RC_{\bar u_i b S^-_k}^RC_{S_l S^-_j S^-_k}G_1(x_{u_i},x_{S^-_j},x_{S^-_k})\nonumber\\
	&&\qquad\qquad\qquad (C_{\mu^-S_l\mu^+}^L+C_{\mu^-S_l\mu^+}^R)\nonumber\\
	&&\qquad\qquad\qquad+\sum_{\chi_i,U_j,U_k,S_l}\frac{1}{2(m_b^2-m^2_{S_l})}m_{\chi_i}C_{\bar s \chi_i U_j}^RC_{\chi_i b U_k}^RC_{S_l U_j U_k}G_1(x_{\chi_i},x_{U_j},x_{U_k})\nonumber\\
	&&\qquad\qquad\qquad (C_{\mu^-S_l\mu^+}^L+C_{\mu^-S_l\mu^+}^R),\nonumber\\
	&&C_{_{p,NP}}^{(2)}(\mu_{_{\rm EW}})=\sum_{u_i,S^-_j,S^-_k,S_l}\frac{1}{2(m_b^2-m^2_{S_l})}m_{u_i}C_{\bar s u_i S^-_j}^RC_{\bar u_i b S^-_k}^RC_{S_l S^-_j S^-_k}G_1(x_{u_i},x_{S^-_j},x_{S^-_k})\nonumber\\
	&&\qquad\qquad\qquad (-C_{\mu^-S_l\mu^+}^L+C_{\mu^-S_l\mu^+}^R)\nonumber\\
	&&\qquad\qquad\qquad+\sum_{\chi_i,U_j,U_k,S_l}\frac{1}{2(m_b^2-m^2_{S_l})}m_{\chi_i}C_{\bar s \chi_i U_j}^RC_{\chi_i b U_k}^RC_{S_l U_j U_k}G_1(x_{\chi_i},x_{U_j},x_{U_k})\nonumber\\
	&&\qquad\qquad\qquad (-C_{\mu^-S_l\mu^+}^L+C_{\mu^-S_l\mu^+}^R),
\end{eqnarray}
\begin{eqnarray}
	&&C_{_{S,NP}}^{(3)}(\mu_{_{\rm EW}})=\sum_{u_i,S^-_k,S_l}\frac{-C_{W^- S_l S^-_k}}{2(m_b^2-m^2_{S_l})}\Big[C_{\bar s W^- u_i}^LC_{\bar u_i S^-_k b}^RG_2(x_{u_i},1,x_{S^-_k})-2m_bm_{u_i}C_{\bar s W^- u_i}^L\nonumber\\
	&&\qquad\qquad\qquad C_{\bar u_i S^-_k b}^LG_1(x_{u_i},1,x_{S^-_k})\Big](C_{\mu^-S_l\mu^+}^L+C_{\mu^-S_l\mu^+}^R),\nonumber\\
	&&C_{_{P,NP}}^{(3)}(\mu_{_{\rm EW}})=\sum_{u_i,S^-_k,S_l}\frac{-C_{W^- S_l S^-_k}}{2(m_b^2-m^2_{S_l})}\Big[C_{\bar s W^- u_i}^LC_{\bar u_i S^-_k b}^RG_2(x_{u_i},1,x_{S^-_k})-2m_bm_{u_i}C_{\bar s W^- u_i}^L\nonumber\\
	&&\qquad\qquad\qquad C_{\bar u_i S^-_k b}^LG_1(x_{u_i},1,x_{S^-_k})\Big](-C_{\mu^-S_l\mu^+}^L+C_{\mu^-S_l\mu^+}^R),
\end{eqnarray}
\begin{eqnarray}
	&&C_{_{S,NP}}^{(4)}(\mu_{_{\rm EW}})=\sum_{u_i,S^-_j,S_l}\frac{-C_{W^- S_l S^-_j}}{2(m_b^2-m^2_{S_l})}C_{\bar s S^-_j u_i}^RC_{\bar u_i W^- b}^RG_2(x_{u_i},x_{S^-_j},1)(C_{\mu^-S_l\mu^+}^L+C_{\mu^-S_l\mu^+}^R),\nonumber\\
	&&C_{_{S,NP}}^{(4)}(\mu_{_{\rm EW}})=\sum_{u_i,S^-_j,S_l}\frac{-C_{W^- S_l S^-_j}}{2(m_b^2-m^2_{S_l})}C_{\bar s S^-_j u_i}^RC_{\bar u_i W^- b}^RG_2(x_{u_i},x_{S^-_j},1)(-C_{\mu^-S_l\mu^+}^L+C_{\mu^-S_l\mu^+}^R),\nonumber\\
\end{eqnarray}
\begin{eqnarray}
	&&C_{_{9,NP}}^{(5)}(\mu_{_{\rm EW}})=\sum_{_{U_i,\chi_j,\chi_k,V
	}}\frac{C_{\mu^-V\mu^+}^L+C_{\mu^-V\mu^+}^R}{-2(m_b^2-m_V^2)}\Big[-\frac{1}{2}C_{U_i\bar s \chi_j}^R C_{\chi_j V \chi_k}^R C_{\chi_k s U_i}^L G_2(x_{U_i},x_{\chi_j},x_{\chi_k})\nonumber\\
	&&\qquad\qquad\qquad+m_{\chi_j}m_{\chi_k} C_{U_i\bar s \chi_j}^R C_{\chi_j V \chi_k}^L C_{\chi_k s U_i}^L G_1(x_{U_i},x_{\chi_j},x_{\chi_k})\Big]\nonumber\\
	&&\qquad\qquad\qquad+\sum_{_{S^-_i,u_j,u_k,V
	}}\frac{C_{\mu^-V\mu^+}^L+C_{\mu^-V\mu^+}^R}{-2(m_b^2-m_V^2)}\Big[-\frac{1}{2}C_{S^-_i\bar s u_j}^R C_{\bar u_j V u_k}^R C_{u_k s S^-_i}^L G_2(x_{S^-_i},x_{u_j},x_{u_k})\nonumber\\
	&&\qquad\qquad\qquad+m_{u_j}m_{u_k} C_{S^-_i\bar s u_j}^R C_{\bar u_j V u_k}^L C_{\bar u_k s S^-_i}^L G_1(x_{S^-_i},x_{u_j},x_{u_k})\Big],\nonumber\\
	&&C_{_{10,NP}}^{(5)}(\mu_{_{\rm EW}})=\sum_{_{U_i,\chi_j,\chi_k,V
	}}\frac{-C_{\mu^-V\mu^+}^L+C_{\mu^-V\mu^+}^R}{-2(m_b^2-m_V^2)}\Big[-\frac{1}{2}C_{U_i\bar s \chi_j}^R C_{\chi_j V \chi_k}^R C_{\chi_k s U_i}^L G_2(x_{U_i},x_{\chi_j},x_{\chi_k})\nonumber\\
	&&\qquad\qquad\qquad+m_{\chi_j}m_{\chi_k} C_{U_i\bar s \chi_j}^R C_{\chi_j V \chi_k}^L C_{\chi_k s U_i}^L G_1(x_{U_i},x_{\chi_j},x_{\chi_k})\Big]\nonumber\\
	&&\qquad\qquad\qquad+\sum_{_{\tilde S^-_i,u_j,u_k,V
	}}\frac{-C_{\mu^-V\mu^+}^L+C_{\mu^-V\mu^+}^R}{-2(m_b^2-m_V^2)}\Big[-\frac{1}{2}C_{S^-_i\bar s u_j}^R C_{\bar u_j V u_k}^R C_{u_k s S^-_i}^L G_2(x_{S^-_i},x_{u_j},x_{u_k})\nonumber\\
	&&\qquad\qquad\qquad+m_{u_j}m_{u_k} C_{S^-_i\bar s u_j}^R C_{\bar u_j V u_k}^L C_{\bar u_k s S^-_i}^L G_1(x_{S^-_i},x_{u_j},x_{u_k})\Big],
\end{eqnarray}
\begin{eqnarray}
	&&C_{_{9,NP}}^{(6)}(\mu_{_{\rm EW}})=\sum_{_{u_i,S^-_j,S^-_k,V
	}}\frac{C_{\mu^-V\mu^+}^L+C_{\mu^-V\mu^+}^R}{4(m_b^2-m_V^2)}C_{\bar s u_i S^-_j}^RC_{\bar u_i b S^-_k}^LC_{V S^-_j S^-_k}G_2(x_{u_i},x_{S^-_j},x_{S^-_k})\nonumber\\
	&&\qquad\qquad\qquad+\sum_{_{\chi_i,U_j,U_k,V
	}}\frac{C_{\mu^-V\mu^+}^L+C_{\mu^-V\mu^+}^R}{4(m_b^2-m_V^2)}C_{\bar s \chi_i U_j}^RC_{\chi_i b U_k}^LC_{V U_j U_k}G_2(x_{\chi_i},x_{U_j},x_{U_k}),\nonumber\\
	&&C_{_{10,NP}}^{(6)}(\mu_{_{\rm EW}})=\sum_{_{u_i,S^-_j,S^-_k,V
	}}\frac{-C_{\mu^-V\mu^+}^L+C_{\mu^-V\mu^+}^R}{4(m_b^2-m_V^2)}C_{\bar s u_i S^-_j}^RC_{\bar u_i b S^-_k}^LC_{V S^-_j S^-_k}G_2(x_{u_i},x_{S^-_j},x_{S^-_k})\nonumber\\
	&&\qquad\qquad\qquad+\sum_{_{\chi_i,U_j,U_k,V
	}}\frac{-C_{\mu^-V\mu^+}^L+C_{\mu^-V\mu^+}^R}{4(m_b^2-m_V^2)}C_{\bar s \chi_i U_j}^RC_{\chi_i b U_k}^LC_{V U_j U_k}G_2(x_{\chi_i},x_{U_j},x_{U_k}),\nonumber\\
\end{eqnarray}
\begin{eqnarray}
	&&C_{_{S,NP}}^{(6)}(\mu_{_{\rm EW}})=\sum_{_{u_i,S^-_j,S^-_k,V
	}}\frac{C_{\mu^-V\mu^+}^L+C_{\mu^-V\mu^+}^R}{-2(m_b^2-m_V^2)}m_bm_{u_i}C_{\bar s u_i S^-_j}^RC_{\bar u_i b S^-_k}^RC_{V S^-_j S^-_k}G_1(x_{u_i},x_{S^-_j},x_{S^-_k})\nonumber\\
	&&\qquad\qquad\qquad+\sum_{_{\chi_i,U_j,U_k,V
	}}\frac{C_{\mu^-V\mu^+}^L+C_{\mu^-V\mu^+}^R}{-2(m_b^2-m_V^2)}m_bm_{\chi_i}C_{\bar s \chi_i U_j}^RC_{\chi_i b U_k}^RC_{V U_j U_k}G_1(x_{\chi_i},x_{U_j},x_{U_k}),\nonumber\\
	&&C_{_{P,NP}}^{(6)}(\mu_{_{\rm EW}})=\sum_{_{u_i,S^-_j,S^-_k,V
	}}\frac{C_{\mu^-V\mu^+}^L-C_{\mu^-V\mu^+}^R}{-2(m_b^2-m_V^2)}m_bm_{u_i}C_{\bar s u_i S^-_j}^RC_{\bar u_i b S^-_k}^RC_{V S^-_j S^-_k}G_1(x_{u_i},x_{S^-_j},x_{S^-_k})\nonumber\\
	&&\qquad\qquad\qquad+\sum_{_{\chi_i,U_j,U_k,V
	}}\frac{C_{\mu^-V\mu^+}^L-C_{\mu^-V\mu^+}^R}{-2(m_b^2-m_V^2)}m_bm_{\chi_i}C_{\bar s \chi_i U_j}^RC_{\chi_i b U_k}^RC_{V U_j U_k}G_1(x_{\chi_i},x_{U_j},x_{U_k}),\nonumber\\
\end{eqnarray}
\begin{eqnarray}
	&&C_{_{9,NP}}^{(7)}(\mu_{_{\rm EW}})=\sum_{_{u_i,S^-_k,V
	}}\frac{C_{\mu^-V\mu^+}^L+C_{\mu^-V\mu^+}^R}{2(m_b^2-m_V^2)}m_{u_i}C_{\bar s u_i W^-}^LC_{\bar u_i b S^-_k}^LC_{V W^- S^-_k}G_1(x_{u_i},x_W,x_{S^-_k}),\nonumber\\
	&&C_{_{10,NP}}^{(7)}(\mu_{_{\rm EW}})=\sum_{_{u_i,S^-_k,V
	}}\frac{-C_{\mu^-V\mu^+}^L+C_{\mu^-V\mu^+}^R}{2(m_b^2-m_V^2)}m_{u_i}C_{\bar s u_i W^-}^LC_{\bar u_i b S^-_k}^LC_{V W^- S^-_k}G_1(x_{u_i},x_W,x_{S^-_k}),\nonumber\\
\end{eqnarray}
\begin{eqnarray}
	&&C_{_{9,NP}}^{(8)}(\mu_{_{\rm EW}})=\sum_{_{u_i,S^-_j,V
	}}\frac{C_{\mu^-V\mu^+}^L+C_{\mu^-V\mu^+}^R}{2(m_b^2-m_V^2)}m_{u_i}C_{\bar s u_i S^-_j}^RC_{\bar u_i b W^-}^LC_{V W^- S^-_k}G_1(x_{u_i},x_{S^-_j},x_W),\nonumber\\
	&&C_{_{10,NP}}^{(8)}(\mu_{_{\rm EW}})=\sum_{_{u_i,S^-_j,V
	}}\frac{-C_{\mu^-V\mu^+}^L+C_{\mu^-V\mu^+}^R}{2(m_b^2-m_V^2)}m_{u_i}C_{\bar s u_i S^-_j}^RC_{\bar u_i b W^-}^LC_{V W^- S^-_k}G_1(x_{u_i},x_{S^-_j},x_W),\nonumber\\
\end{eqnarray}
\begin{eqnarray}
	&&C_{_{9,NP}}^{(9)}(\mu_{_{\rm EW}})=\sum_{_{U_i,\chi_j,\chi_k,S_l
	}}-\frac{1}{8}C_{\bar s U_i \chi_j}^RC_{\bar\chi_j \mu^+ S_l}^L(C_{\bar\mu^-S_l\chi_k}^LC_{\bar \chi_k U_i b}^R+C_{\bar\mu^-S_l\chi_k}^RC_{\bar \chi_k U_i b}^L)\nonumber\\
	&&\qquad\qquad\qquad G_4(x_{U_i},x_{\chi_j},x_{\chi_k},x_{\chi^0_l}),\nonumber\\
	&&C_{_{10,NP}}^{(9)}(\mu_{_{\rm EW}})=\sum_{_{U_i,\chi_j,\chi_k,S_l
	}}-\frac{1}{8}C_{\bar s U_i \chi_j}^RC_{\bar\chi_j \mu^+ S_l}^L(C_{\bar\mu^-S_l\chi_k}^LC_{\bar \chi_k U_i b}^R-C_{\bar\mu^-S_l\chi_k}^RC_{\bar \chi_k U_i b}^L)\nonumber\\
	&&\qquad\qquad\qquad G_4(x_{U_i},x_{\chi_j},x_{\chi_k},x_{\chi^0_l}),\nonumber\\
\end{eqnarray}
\begin{eqnarray}
	&&C_{_{S,NP}}^{(9)}(\mu_{_{\rm EW}})=\sum_{_{U_i,\chi_j,\chi_k,S_l
	}}-\frac{1}{2}m_{\chi_j}m_{\chi_k}C_{\bar s U_i \chi_j}^RC_{\bar\chi_j \mu^+ S_l}^R(C_{\bar\mu^-S_l\chi_k}^LC_{\bar \chi_k U_i b}^L+C_{\bar\mu^-S_l\chi_k}^RC_{\bar \chi_k U_i b}^R)\nonumber\\
	&&\qquad\qquad\qquad G_3(x_{U_i},x_{\chi_j},x_{\chi_k},x_{\chi^0_l}),\nonumber\\
	&&C_{_{P,NP}}^{(9)}(\mu_{_{\rm EW}})=\sum_{_{U_i,\chi_j,\chi_k,S_l
	}}-\frac{1}{2}m_{\chi_j}m_{\chi_k}C_{\bar s U_i \chi_j}^RC_{\bar\chi_j \mu^+ S_l}^R(-C_{\bar\mu^-S_l\chi_k}^LC_{\bar \chi_k U_i b}^L+C_{\bar\mu^-S_l\chi_k}^RC_{\bar \chi_k U_i b}^R)\nonumber\\
	&&\qquad\qquad\qquad G_3(x_{U_i},x_{\chi_j},x_{\chi_k},x_{\chi^0_l}),
\end{eqnarray}
\begin{eqnarray}
	&&C_{_{9,NP}}^{(10)}(\mu_{_{\rm EW}})=\sum_{_{u_i,S^-_j,S^-_k,S_l}}\frac{1}{8}C_{\bar s u_i S^-_j}^RC_{\bar u_i b S^-_k}^L(C_{\bar\mu^-S^-_k\chi^0_l}^LC_{\bar S_l S^-_j \mu^+}^R+C_{\bar\mu^-S^-_k\chi^0_l}^RC_{\bar S_l S^-_j \mu^+}^L)\nonumber\\
	&&\qquad\qquad\qquad G_4(x_{u_i},x_{S^-_j},x_{S^-_k},x_{\chi^0_l}),\nonumber\\
	&&C_{_{10,NP}}^{(10)}(\mu_{_{\rm EW}})=\sum_{_{u_i,S^-_j,S^-_k,S_l}}\frac{1}{8}C_{\bar s u_i S^-_j}^RC_{\bar u_i b S^-_k}^L(C_{\bar\mu^-S^-_k\chi^0_l}^LC_{\bar S_l S^-_j \mu^+}^R-C_{\bar\mu^-S^-_k\chi^0_l}^RC_{\bar S_l S^-_j \mu^+}^L)\nonumber\\
	&&\qquad\qquad\qquad G_4(x_{u_i},x_{S^-_j},x_{S^-_k},x_{\chi^0_l}),
\end{eqnarray}
\begin{eqnarray}
	&&C_{_{S,NP}}^{(11)}(\mu_{_{\rm EW}})=-\sum_{_{u_i,S^-_j,\chi^0_l}}\frac{1}{2}C_{\bar s u_i S^-_j}^RC_{\bar u_i b W^-}^RC_{\bar\mu^-W^-\chi^0_l}^RC_{\bar S_l S^-_j \mu^+}^L G_4(x_{u_i},x_{S^-_j},x_W,x_{\chi^0_l}),\nonumber\\
	&&C_{_{P,NP}}^{(11)}(\mu_{_{\rm EW}})=-\sum_{_{u_i,S^-_j,\chi^0_l}}\frac{1}{2}C_{\bar s u_i S^-_j}^RC_{\bar u_i b W^-}^RC_{\bar\mu^-W^-\chi^0_l}^RC_{\bar S_l S^-_j \mu^+}^L G_4(x_{u_i},x_{S^-_j},x_W,x_{\chi^0_l}),
\end{eqnarray}
where $C_{abc}^{L,R}$ denotes the constant parts of the interaction vertex about $abc$, and $a, b, c$ denote the interactional particles, $V$ denotes photon $\gamma$ and $Z$ boson. $L$ and $R$ in superscript denote the left-hand part and right-hand part.

Denoting $x_i=\frac{m_i^2}{m_W^2}$, the concrete expressions for $G_k(k=1,...,4)$ can be given as:
\begin{eqnarray}
	&&G_1(x_1,x_2,x_3)=\frac{-1}{m_W^2}\Big[\frac{x_1{\rm ln} x_1}{(x_2-x_1)(x_3-x_1)}+\frac{x_2{\rm ln} x_2}{(x_1-x_2)(x_3-x_2)}+\frac{x_3{\rm ln} x_3}{(x_1-x_3)(x_2-x_3)}\Big],\nonumber\\
	&&G_2(x_1,x_2,x_3)=-\frac{x_1^2{\rm ln} x_1}{(x_2-x_1)(x_3-x_1)}-\frac{x_2^2{\rm ln} x_2}{(x_1-x_2)(x_3-x_2)}-\frac{x_3^2{\rm ln} x_3}{(x_1-x_3)(x_2-x_3)},\nonumber\\
	&&G_3(x_1,x_2,x_3,x_4)=\frac{1}{m_W^4}\Big[\frac{x_1{\rm ln} x_1}{(x_2-x_1)(x_3-x_1)(x_4-x_1)}+\frac{x_2{\rm ln} x_2}{(x_1-x_2)(x_3-x_2)(x_4-x_2)}+\nonumber\\
	&&\qquad\qquad\qquad\qquad\frac{x_3{\rm ln} x_3}{(x_1-x_3)(x_2-x_3)(x_4-x_3)}\frac{x_4{\rm ln} x_4}{(x_1-x_4)(x_2-x_4)(x_3-x_4)}\Big],\nonumber\\
	&&G_4(x_1,x_2,x_3,x_4)=\frac{1}{m_W^2}\Big[\frac{x_1^2{\rm ln} x_1}{(x_2-x_1)(x_3-x_1)(x_4-x_1)}+\frac{x_2^2{\rm ln} x_2}{(x_1-x_2)(x_3-x_2)(x_4-x_2)}+\nonumber\\
	&&\qquad\qquad\qquad\qquad\frac{x_3^2{\rm ln} x_3}{(x_1-x_3)(x_2-x_3)(x_4-x_3)}\frac{x_4^2{\rm ln} x_4}{(x_1-x_4)(x_2-x_4)(x_3-x_4)}\Big].\nonumber\\
\end{eqnarray}

\end{document}